\documentclass[reprint,showpacs,preprintnumbers,amsmath,amssymb,twocolumn]{revtex4}

\usepackage{graphicx}
\graphicspath{ }

\usepackage{dcolumn}

\usepackage{bm}

\usepackage{relsize}                    
\usepackage{amsmath}

\usepackage{amssymb}

\usepackage{url}

\usepackage{caption}		

\def \bea{\begin{eqnarray}}

 \def\eea{\end{eqnarray}}

\begin{document}


\title{Configuration interaction matrix elements for the quantum Hall effect}


\author{R. E. Wooten}
\altaffiliation{Department of Physics and Astronomy, Purdue University, West Lafayette IN 47907-2036}

\author{J. H. Macek}


\affiliation{Department of Physics and Astronomy, University of Tennessee, Knoxville TN 37996-1501 }

\date{\today}


\begin{abstract}
We derive analytic expressions for the two-body matrix elements in finite spherical quantum Hall systems in terms of a general scalar interaction expressed as a sum over Legendre polynomials, and we derive the corresponding pair pseudopotentials from these matrix elements. The relationship between the effective spatial potential and the pseudopotential is one-to-one in this framework, and we show how any complete model pseudopotential can be analytically inverted to give a unique corresponding spatial potential. As an example, we find the spatial potential that produces a harmonic pseudopotential and verify that it fails to break the angular momentum degeneracy of the many-body quantum Hall system. We also include additional examples to demonstrate the use of the inversion technique.
\end{abstract}


\maketitle

\section{\label{sec:intro} Introduction}

The quantization of the quantum Hall system into highly degenerate Landau levels is both very well understood and experimentally robust \cite{Klitzing1980}, and electrons in a partially filled Landau level (LL) constitute the ultimate example of a highly-interacting system. For very strong magnetic fields, the Coulomb interaction and the magnetic confinement are the dominant factors dictating the collective electronic behavior in the system, and, in comparison to many other strongly interacting systems, the quantum Hall system is relatively simple to model. But even though the model is not exceptionally complicated, numerical calculations give strong enough agreement with experimental results that the calculations are treated as numerical experiments.

The quantum Hall system consists of electrons confined to a very thin metallic plane in the presence of a perpendicular, strong magnetic field. For a system with $N$ electrons all in the same Landau level, the energy spectrum can be found by diagonalizing the effective two-dimensional, two-body spatial interaction $V(r_{ij})$. The spatial interaction is usually assumed to be the Coulomb interaction, but it may also be modified by the z-direction quantum well \cite{MacDonald1984, Zhang1986, Park1998PRL, Peterson2008} or Landau level mixing \cite{Yoshioka1984, Rezayi1990, Wojs2006, Bishara2009, Simon2013, Sodemann2013, Peterson2013, Wooten2013}. Numerical studies (e.g. \cite{Park1998PRB, Toke2006}) are also performed using effective spatial potentials that nearly reproduce the desired pseudopotential, the interaction energy of a pair of particles as a function their combined angular momentum \cite{Haldane1983,Prange1987}.

While it seems most natural to perform numerical calculations using the same geometry as the experimental system, calculations are frequently performed in two other geometric configurations: on the surface of a sphere, and on the surface of a torus. In this paper, we will derive expressions for the pair interactions and the pseudopotential in the spherical geometry, which is known as the Haldane sphere. The planar and spherical systems are linked by a stereographic mapping \cite{Fano1986}, and the pseudopotential on the sphere is equivalent the planar pseudopotential (they are numerically indistinguishable in the limit of the infinite sphere). Although the two system topologies are locally equivalent, the Haldane sphere is particularly useful because it lacks boundary conditions and because calculations in the spherical system can take full advantage of the well-developed mechanics of angular momentum theory.  

We will begin in section \ref{sec:II} of this paper by presenting the interaction matrix elements for particles in any Landau level in the Haldane sphere model for a general two-body scalar interaction, $V(r_{12})$, expressed as a sum over Legendre polynomials. The $k^{th}$ Legendre polynomial in the sum will by weighted by a coefficient, $V_k$, and the complete infinite set of $V_k$ coefficients defines the potential, and can be used to model finite width effects. Although an infinite sum is necessary to completely define $V(r)$ in general, the two-body matrix elements are given by finite sums over angular momentum coupling coefficients times $V_k$ coefficients, and Legendre polynomials beyond a certain index, $k_{max}$, do not affect dynamics of a given finite system. As a result, only a finite set of $V_k$ coefficients are necessary to completely express the effective potentials for finite systems. 

Section \ref{sec:III} will present the related analytic expression for the Haldane pseudopotential in terms of these same $V_k$ parameters.  While the properties of the monopole harmonics, and the very closely related spin-weighted spherical harmonic and Wigner D functions, are well-described in the literature \cite{Wu1976, Wu1977, Scanio1977, Dray1985, Dray1986}, direct expressions for the two-body matrix elements are uncommon (\cite{Fano1986} gives the matrix elements in terms of the pseudopotential, but only for the lowest LL; likewise, \cite{Jain2007} sets up the two-body matrix element integrals in much the same manner as this paper, but leaves their evaluation to the reader).  Meanwhile, spherical pseudopotentials are generally presented without derivation, as in \cite{Ginocchio1993, Toke2006} or are given for only the lowest LL. We have included the details of both derivations as appendices, primarily for archival interest.

In section \ref{sec:IV}, we will show that, for the restricted functional space of a single Landau level in a finite spherical system, there is a one-to-one functional relationship between the effective spatial potential of section \ref{sec:II} and the finite single LL pseudopotential of section \ref{sec:III}. In other words, for a single finite LL defined by a shell angular momentum $l$, the effective spatial potential is defined by the first $2l+1$ terms in the Legendre expansion while the pseudopotential is comprised of $2l+1$ terms. The relationship between the effective potential and the pseudopotential can be represented as an invertible matrix equation. We demonstrate how this matrix equation can be analytically inverted to give the unique effective spatial potential as a function of the terms of the pseudopotential, and will discuss limitations of the method.

In section \ref{sec:V}, as a demonstration of the inversion technique presented in section \ref{sec:IV}, we take a general Harmonic pair pseudopotential, a pseudopotential of the form $V_H(L) = a + b L(L+1)$, and analytically invert it to recover the effective spatial potential that produces the Harmonic pseudopotential. This particular pseudopotential is interesting because, as proven by W\'ojs and Quinn \cite{Wojs1998,Quinn2000} it does not split the energy degeneracy of $N$-particle states with the same total angular momentum $L$, and serves as a boundary between systems exhibiting Laughlin or non-Laughlin correlations.  This section will verify the theorem by Quinn and W\'ojs, will connect their work to the work of Johnson \cite{Johnson1991, Johnson2002}, and will provide additional insight into why the theorem is valid.

And finally in section \ref{sec:VI}, we will use several examples to demonstrate the inversion procedure. We will use this section primarily to draw attention to issues that may arise during numerical calculations and to show how to get a useful spatial potential for reasonable model pseudopotentials.

\section{\label{sec:II}Matrix elements}

Before defining the two-particle interaction matrix elements, we will first briefly review the single-particle solutions on the Haldane sphere. The model consists of a spherical metallic shell centered around a Dirac magnetic monopole of strength $2Q\phi_0$, where $\phi_0 = hc/e$ is the fundamental magnetic flux quantum, and $Q$ is an integer or half-integer. The resulting uniform magnetic field at the sphere's surface is $\mathbf{B} = 2Q\phi_0/4\pi R_{s}^2\hat{R}$, where $R_s$ is the radius of the sphere. The resulting single-particle Hamiltonian is given by 
\begin{equation}
\hat{H}_0 = \frac{1}{2mR_s^2} \left(  \hat{\mathbf{l}}^2 - Q^2 \right) \label{eq2.1}
\end{equation}
Here, $\hat{\mathbf{l}}$ is the standard angular momentum operator. The single particle solutions to the Schr$\ddot{o}$dinger equation $\hat{H}_0 \psi = E \psi$ are known as the monopole harmonics, $Y_{Q,l,m}$ or $|Q,l,m\rangle$. They are defined in the papers by Wu and Yang \cite{Wu1976,Wu1977} in terms of the rotation functions defined in reference \cite{Edmonds1996}. The quantum numbers $l$ and $m$ are the standard shell-angular momentum and azimuthal quantum numbers, respectively, with
\begin{align}
\hat{l}^2 |Q,l,m\rangle = \hbar^2 l(l+1)|Q,l,m\rangle \nonumber \\
\hat{l}_z |Q, l, m\rangle = \hbar m |Q, l, m\rangle \nonumber \\
\hat{H}_0 |Q, l, m \rangle = \frac{\hbar \omega_c}{2Q}\left[l(l+1) - Q^2\right]|Q, l, m\rangle,  \label{eq2.2}
\end{align}
where $\omega_c = eB/mc$ is the cyclotron frequency (in CGS units). The $g^{th}$ Landau level is defined by the shell angular momentum $l = Q+g$ with $g \geq 0$, and each energy level is $(2l+1)$-fold degenerate in the azimuthal quantum number, $m$.

Particles, typically electrons, are assumed to interact in this system according to a general two-body scalar potential, $V(|\mathbf{r}_1-\mathbf{r_2}|) = V(r_{12})$, such as the Coulomb potential. In the ideal quantum Hall system, the metallic sheet is infinitely thin, so that in the Haldane model, $|\mathbf{r}_1| = r_1 = r_2 = |\mathbf{r}_2|$, but in reality, the sheet has a finite thickness, and the particles are trapped by a narrow potential well. The full two-body matrix element consists of a radial and an angular part.  For a thin sheet, all particles lie close enough to the ideal spherical shell with radius $R_s$ that the magnetic field remains constant for all particles in the model, and the angular solutions are the monopole harmonics sharing the same monopole strength, $2Q$. 

As long as the thin-sheet approximation is good, we can express any scalar central interaction, $V(r_{12})$, as a function of the particles' angular separation, $\theta_{12}$ on the Haldane sphere. The potential can then be written as an infinite sum over Legendre polynomials, where the $k^{th}$ is weighted by a coefficient $V_k$:
\begin{equation}
V(\theta_{12}) = \left( \frac{e^2}{4 \pi \epsilon \lambda_0}\right)\frac{1}{\sqrt{Q}} \sum_{k=0}^\infty V_k P_k(\cos \theta_{12}). \label{eq2.4}
\end{equation} 
The set of $V_k$ coefficients can be used to represent any arbitrary scalar potential, and can be used to incorporate finite well width effects. Here, we have chosen to have the $V_k$ coefficients be unit-less, and so have factored out $e^2/(4\pi \epsilon R_s) = e^2/ (4 \pi \epsilon \lambda_0 \sqrt{Q})$, where $\epsilon$ is the dielectric constant of the material, and $\lambda_0 = \sqrt{\hbar c /eB}$ is the magnetic length. We have also absorbed the factor of $(2k+1)$ of \cite{Edmonds1996} into each of the $V_k$ coefficients of equation \eqref{eq2.4}. The most common choice for the $V_k$ is $V_k = 1$ for all values of $k$, which gives the pure Coulomb potential. 

Using these constants to define the potential, the full two-body matrix element in units of the Coulomb energy, $e^2/ (4 \pi \epsilon \lambda_0$), is
\begin{align}
&\langle  l_1', m_1';   l_2', m_2' | V(r_{12})|  l_1, m_1;   l_2, m_2 \rangle =  \nonumber \\
 & \frac{1}{\sqrt{Q}}  \sum_{k=0}^\infty V_k  \langle l_1', m_1' ;  l_2', m_2'  | P_k(\cos \theta_{12}) | l_1,m_1 ; l_2, m_2  \rangle.  \label{eq2.5}
\end{align}
The monopole strength quantum number, $Q$, is the same for all states because all particles are embedded in the same Haldane surface with the same constant magnetic field, and it has been dropped from the state vectors here for brevity. Note that we have made no assumptions about the symmetry of the particles in this system: the state vector $| l_1, m_1; l_2, m_2 \rangle$ is neither explicitly symmetric nor antisymmetric, and for the full calculation of matrix elements, the wave function symmetry must also be considered. Since we have not imposed any symmetry considerations, our derived equations can be applied to indistinguishable fermions and bosons and also to distinguishable particles (for example, electron-hole pairs).

Evaluating equation \eqref{eq2.5} requires standard angular momentum theory, but special care must be taken because the monopole harmonics behave similarly, but not identically to the more familiar spherical harmonics. We show the details of the derivation in appendix \ref{appendix:i}. The resulting two-body matrix element is proportional to a finite sum over Wigner 3J symbols times the Legendre polynomial coefficients, $V_k$:
\begin{align}
\langle l_1', m_1';  l_2', m_2' | V(&r_{12})| l_1, m_1; l_2, m_2 \rangle =\nonumber \\
 \frac{\bar{l} (-1)^{\eta} }{ \sqrt{Q}}  \sum_{k = 0}^{k_{max}}  &V_k   \left[
 \begin{pmatrix}
l_1' & k & l_1 \\
-m_1' & -m & m_1
\end{pmatrix}
\begin{pmatrix}
l_1' & k & l_1 \\
-Q & 0 & Q
\end{pmatrix} \right.    \nonumber \\
&\times \left.
\begin{pmatrix}
l_2' & k & l_2\\
-m_2' & m & m_2
\end{pmatrix}
\begin{pmatrix}
l_2' & k & l_2 \\
-Q & 0 & Q
\end{pmatrix} \right].     \label{eq2.10}
\end{align}  
The value of $\bar{l}$ is given by $\bar{l} = \left[(2l_1'+1)(2l_1+1)(2l_2'+1)(2l_2+1)\right]^{\frac{1}{2}}$, and $\eta$ is given by $\eta = 2Q + m2' + m_1 + l_1'+ l_2'+ l_1 + l_2 $. The matrix element is given in units of the Coulomb energy, $e^2/4\pi\epsilon\lambda_0$, where $\lambda_0 = \sqrt{\hbar c/eB}$. In addition, because the 3J symbols are non-zero only when the upper three indices satisfy the triangle condition, $|l_1' - l_1| \leq k \leq l_1+l_1'$, the infinite sum over $k$ in equation \eqref{eq2.5} terminates after $k$ becomes too large. The triangle inequality requirement reduces the infinite sum over $k$ to a finite sum up to $k_{max}$, where $k_{max}$ is the largest integer that satisfies both $k_{max} \leq l_1+l_1'$ and $k_{max} \leq l_2+l_2'$. 

Although the potential $V(\theta_{12})$ is defined in equation \eqref{eq2.4} as an infinite sum over Legendre polynomials, only a finite number of terms of that sum actually contribute to configuration interaction matrix element calculations. The effective potential of any finite system, then, is the sum over the finite set of Legendre polynomials that actually affect the particle dynamics, 
\begin{equation}
V_{eff}(\theta_{12}) = \left( \frac{e^2}{4 \pi \epsilon \lambda_0}\right)\frac{1}{\sqrt{Q}} \sum_{k=0}^{k_{max}}V_k P_k(\cos \theta_{12}), \label{eq2.11}
\end{equation}
where $k_{max}$ is determined by the triangle inequalities based on $Q$ and the LLs of the specific system. 

\begin{figure}
                \centering
                \includegraphics[width=.5\textwidth]{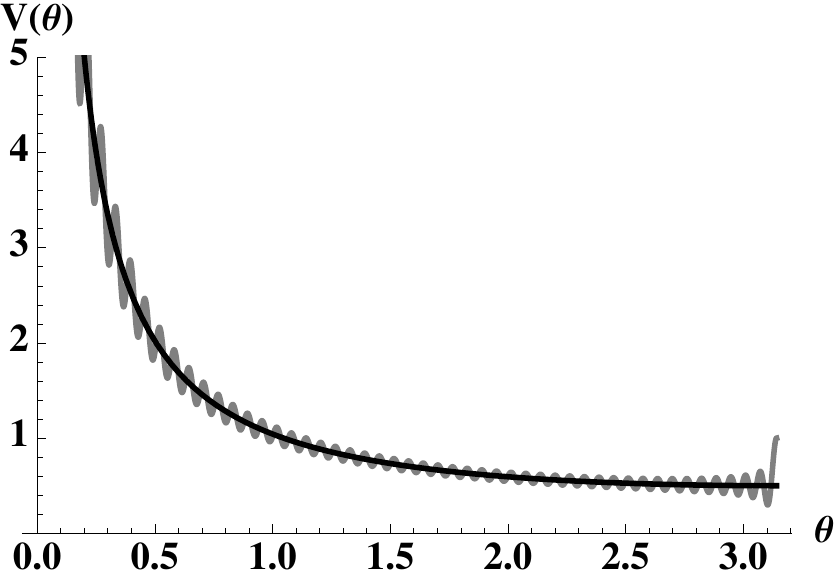}
                \label{fig1}
\caption{The effective Coulomb potential for a finite system as a function of the angular particle separation, $\theta$, using equation \eqref{eq2.11} with $k_{max}=50$, and $V_k = 1$ for all allowed $k$. The genuine Coulomb potential, $V(r) = 1/r$ is shown in black, where r is the chord distance between particles, given in terms of $\theta$ as $r/R = 2 \sin(\theta/2)$, where $R$ is the radius of the sphere. The effective Coulomb potential exhibits ringing in all non-infinite systems, but the minor oscillations do not affect the matrix element calculations for finite systems and are an artifact of performing calculations in a non-infinite system.
}
\end{figure}

The effective potential given by equation \eqref{eq2.11} versus angle or inter-particle distance will often exhibit oscillatory ringing behavior. For example, consider the matrix element calculations of the pure Coulomb potential for electrons confined to the $n^{th}$ Landau level in a finite system defined by a non-infinite shell angular momentum, $l = n + Q$. The true Coulomb potential is given by equation \eqref{eq2.4} with $V_k = 1$ for $0 \leq k < \infty$, but in equation \eqref{eq2.10}, no terms of the sum beyond $k_{max} = 2l$ contributes to the matrix element calculation due to the the orthogonality properties of the angular momentum coupling coefficients. The plot of the effective Coulomb potential for any non-infinite system is apparently oscillatory, and will fail to precisely match the exact Coulomb potential (see figure \ref{fig1}), but this non-matching is irrelevant to the particle dynamics in the finite system.

\section{\label{sec:III}The pair pseudopotential} 

On the Haldane sphere, the pair pseudopotential, $V(L)$, is the interaction energy of a pair of particles as a function of their pair angular momentum, $L$. The pseudopotential contains the same information as the spatial interaction, $V(r_{12})$, and can be used in its place to perform calculations. The pseudopotential is defined on the Haldane sphere by using standard angular momentum coupling to expand the monopole harmonics into the coupled basis,
\begin{align}
|  l_1, m_1;  l_2, m_2 \rangle =& \nonumber \\
 \sum_{L,M}  |  l_1,  l_2; & L, M \rangle   \langle l_1, l_2; L, M |  l_1, m_1;  l_2, m_2 \rangle, \label{eq3.1}
\end{align}
where again, the monopole quantum number $Q$ has been dropped from the state vectors for brevity. Equation \eqref{eq3.1} is analogous to equation (3.5.2) in \cite{Edmonds1996}, except that the angular momentum eigenvectors, $| l_1, l_2; L, M \rangle$, are coupled monopole harmonics instead of coupled spherical harmonics.  The expression $\langle l_1, l_2; L, M | l_1, m_1; l_2, m_2 \rangle$ is equal to an ordinary Clebsch-Gordan coefficient \cite{Wu1976}. Note here that the coupled states are neither explicitly symmetric nor antisymmetric: as in the previous section, particle exchange and wave function symmetry are not included in the derivation. A brief discussion of the symmetric and antisymmetric pair functions has been included in appendix \ref{appendix:ii}.

The evaluation of a matrix element like equation \eqref{eq2.5} can be performed by expanding the pair functions in the coupled basis using \eqref{eq3.1}. The resulting matrix element equation contains the expression 
\begin{align}
\langle  l_1', l_2'; L', M' | V(r_{12}) |  l_1, l_2; L, M \rangle =& \nonumber \\
\frac{1}{\sqrt{Q}} \sum_{k=0}^{\infty} V_k     \langle  l_1', l_2'; L', M' | P_k (\cos &\theta_{12})|   l_1, l_2; L, M \rangle, \label{eq3.2}
\end{align}
which gives the pseudopotential for the $g^{th}$ LL when evaluated for $l_1 = l_2 = l_1' = l_2' = Q+g$. The general expression for the pseudopotential for the unmodified Coulomb potential has been previously given by \cite{Ginocchio1993}, but is presented there without a derivation. We have included the details of the derivation in appendix \ref{appendix:iii}. The pseudopotential for particles in any LL without regard to the particle symmetry is given in terms of the Wigner 3J and 6J symbols by the following equation:
\begin{align}
 \langle l_1', l_2'; L', M' | V(r_{12}) |  l_1, l_2; L, M \rangle =
\delta_{L, L'}  \delta_{M,M'} \frac{\bar{l}(-1)^{\eta}}{\sqrt{Q}}  \nonumber \\
\times \sum_{k=0}^{k_{max}} V_k   
\begin{Bmatrix}
 L & l_2' & l_1' \\
 k & l_1 & l_2
 \end{Bmatrix}
 \begin{pmatrix}
 l_1' & k & l_1 \\
 -Q & 0 & Q
 \end{pmatrix}
 \begin{pmatrix}
 l_2' & k & l_2 \\
 -Q & 0 & Q
 \end{pmatrix}. \label{eq3.8}
\end{align}
Here, $\eta = 2Q+L+2l_1+l_2+l_2'$, $\delta$ is the standard Kronecker delta, and $\bar{l} = \left[(2l_1'+1)(2l_1+1)(2l_2'+1)(2l_2+1)\right]^{\frac{1}{2}}$. The value of $k_{max}$ is determined by the triangle inequalities for the sets $\{ l_1', k, l_1 \}$ and $\{ l_2', k, l_2 \}$ as it was in section \ref{sec:II}. 

Most often, the pseudopotential is desired only for particles all in the same LL, with $l_1 = l_1' = l_2 = l_2' = l$. In this case the pseudopotential without regard to particle symmetry is given by
\begin{align}
V_{Q,l}(L) = & 
\frac{(-1)^{2Q+L}}{\sqrt{Q}} (2l+1)^2 \nonumber \\  
& \times\sum_{k=0}^{2l} V_k  
   \begin{Bmatrix}
 L & l & l \\
 k & l & l
 \end{Bmatrix}
 \begin{pmatrix}
 l & k & l \\
 -Q & 0 & Q
 \end{pmatrix}^2. 	 \label{eq3.9}
\end{align}
Equations \eqref{eq3.8} and \eqref{eq3.9}, like equation \eqref{eq2.10}, are given by finite sums over angular momentum coupling coefficients times the $V_k$ coefficients defining an effective spatial potential in a finite system.  Although $V_{Q,l}(L)$ is a function of $Q$ and $l$, these are constants for a single Landau level in a specific Haldane sphere system, and the pseudopotential on the sphere approaches the pseudopotential on the infinite plane asymptotically as $Q$ approaches infinity (i.e. for an infinite sphere). Since $Q$ and $l$ do not vary in a given problem, they can be dropped as labels for $V(L)$ in \eqref{eq3.9}, assuming they are known from context. More importantly, the pseudopotential is independent of $M$ and diagonal in $L$ and $M$ as a result of the Wigner-Eckart theorem. 

\section{\label{sec:IV}Inverting the Pseudopotential}
Given a complete set of $2l+1$ $V_k$ coefficients defining the effective potential for a single LL according to \eqref{eq2.11} we can evaluate the full pseudopotential $V_{Q,l}(L)$ for particles in the same Landau level. It is also possible to solve the inverse problem: given a complete pseudopotential for a single given LL (that is, the set of all $2l+1$ values of the pseudopotential for $0 \leq L \leq 2l$), we can solve for the set of $2l+1$ $V_k$ coefficients of the effective potential that produce that pseudopotential. This can easily be seen if we rewrite equation \eqref{eq3.9} as a matrix equation,
\begin{equation}
\mathbf{V}_L = \mathbf{M}_{L,k} \mathbf{V}_k  \label{eq4.0}
\end{equation}
where $\mathbf{V}_L$ is a column array of the $2l+1$ pseudopotential values of the $n^{th}$ LL, $\mathbf{V}_k$ is a column array of the $2l+1$ $V_k$ coefficients defining the effective spatial potential, and $\mathbf{M}_{L,K}$ is an invertible $(2l+1)$ by $(2l+1)$ dimensional matrix. From the form of equation \eqref{eq4.0}, it is clear that a unique effective potential can be found for a completely defined pseudopotential simply by multiplying both sides by $\mathbf{M}^{-1}$ from the left. 

Although the $2l+1$ $V_k$ coefficients of the effective potential could be calculated from $V(L)$ by numerically inverting equation \eqref{eq4.0}, the solution to this inverse problem can also be found analytically by using one of the orthogonality properties of the 6J symbols, equation (6.2.9) from reference \cite{Edmonds1996}:
\begin{align}
\sum_j (2j+1)(2j''+1)    \begin{Bmatrix}
j_1 & j_2 & j' \\
 j_3 & j_4 & j
 \end{Bmatrix}
 \begin{Bmatrix}
j_3 & j_2 & j \\
 j_1 & j_4 & j''
 \end{Bmatrix}
  = \delta_{j',j''}	\label{eq4.1}
\end{align} 

We introduce a new dummy variable, $j$, to \eqref{eq3.9} by multiplying both sides by a function of $L$ and $j$ to yield 
\begin{align}
V_{Q,l}(L) &\left[ (-1)^{-L} (2L+1)(2j+1)   
\begin{Bmatrix}
 l & l & j \\
l & l & L
 \end{Bmatrix} 
 \right] = \nonumber \\
&\frac{1}{\sqrt{Q}} (2l+1)^2 (-1)^{2Q} 
\sum_{k=0}^{2l}  \left[ V_k 
\begin{pmatrix}
 l & k & l \\
 -Q & 0 & Q
 \end{pmatrix}^2 \right. \nonumber \\
 & \times (2L+1)(2j+1)
 \begin{Bmatrix}
 l & l & j \\
l & l & L
 \end{Bmatrix}
\left.
 \begin{Bmatrix}
 l & l & L \\
l & l & k
 \end{Bmatrix}
 \right] .     \label{eq4.2}
\end{align}
Taking the sum of both sides of \eqref{eq4.2} over all allowed values of $L$ and applying the orthogonality property of equation \eqref{eq4.1} gives
\begin{align}
\sum_{L=0}^{2l} V_{Q,l}(L) & \left[ (-1)^{-L} (2L+1)(2j+1)   \begin{Bmatrix}
 l & l & j \\
l & l & L
 \end{Bmatrix} 
 \right] = \nonumber \\
 & \frac{1}{\sqrt{Q}} (2l+1)^2 (-1)^{2Q}  \nonumber \\
 & \times \sum_{k=0}^{2l} V_k 
\begin{pmatrix}
 l & k & l \\
 -Q & 0 & Q
 \end{pmatrix}^2
 \delta_{k,j}
 .     \label{eq4.3}
\end{align}
The Kronecker delta in \eqref{eq4.3} collapses the sum over $k$. Replacing the dummy variable $j$ with the more familiar $k$, the $V_k$ coefficients can be calculated directly as a sum over pseudopotential values times vector coupling coefficients,
\begin{align}
V_k &=  \frac{\sqrt{Q}}{(2l+1)^2}(-1)^{2Q}
\begin{pmatrix}
 l & k & l \\
 -Q & 0 & Q
 \end{pmatrix}^{-2} \nonumber \\
&\times \sum_{L = 0}^{2l} 
 V_{Q,l}(L) (-1)^{L} (2L+1)(2k+1) 
  \begin{Bmatrix}
 l & l & k \\
l & l & L
 \end{Bmatrix}. \label{eq4.4}
\end{align}
Since $L$ must be an integer for particle pairs in this system, we have included the minor simplification $(-1)^{-L} = (-1)^L$. It is simple enough to verify that equation \eqref{eq4.4} properly inverts a Coulomb pseudopotential, where $V_k = 1$ for all $k \leq 2l$ and $V_{Q,l}(L)$ is calculated using equation \eqref{eq3.9}.  

If we consider only finite systems and isolated LLs, it is clear from equations \eqref{eq4.0} and \eqref{eq4.4} that the relationship between the effective spatial potential (defined by the truncated, $V_k$-weighted sum of the first $2l+1$ Legendre polynomials) and the pseudopotential (with all $2l+1$ terms, including both even and odd values of $L$) is one-to-one. In other words, a completely defined pseudopotential corresponds to a unique effective potential for finite systems.  

However, in many cases, the complete pseudopotential behavior is not necessary for predicting the dynamics of the system. For example, a pair of spin polarized electrons in the same LL, must have a spatially anti-symmetric wave function, and their pair angular momentum must necessarily follow the rule $2l-L = \mathcal{R} = odd$. The interaction behavior of such an electron pair is unaffected by changes to the even $\mathcal{R}$ pseudopotential terms, and there are any number of pseudopotentials with differing even $\mathcal{R}$ values that produce the same electronic behavior. While the inversion of a completely defined pair pseudopotential gives a unique spatial potential, there are an infinite number of pseudopotentials with identical odd $\mathcal{R}$ values but differing even $\mathcal{R}$ that produce the same spin-polarized electron behavior. Unfortunately, inverting different pseudopotentials with different even $\mathcal{R}$ behavior will yield different spatial potentials, and it is not possible in general to predict any specific form for all spatial potentials with matching values of the odd values of $\mathcal{R}$ only. However, incorporating realistic values of the pseudopotential for the even values of $\mathcal{R}$ for the purpose of finding a realistic spatial potential should not prove too great of a difficulty when modeling electronic systems.

Because inverting the pseudopotential constitutes inverting a two-dimensional matrix, it is difficult to predict in general which types of spatial potentials will yield specific types of pseudopotential, but when it is possible in certain cases to invert a somewhat generalized pseudopotential to yield a class of spatial potentials. Of particular interest is the harmonic pseudopotential, a pseudopotential given by $V(L) = A + BL(L+1)$, where $A$ and $B$ are arbitrary constants.  The harmonic pseudopotential is of particular note because it does not break the angular degeneracy of the system. It also serves as the pseudopotential that separates Laughlin correlated behavior from non-Laughlin correlated behavior \cite{Wojs1998, Quinn2000}. 

\section{\label{sec:V}Harmonic pseudopotentials}

Beginning with an arbitrary harmonic pseudopotential, $V_{Q,l}(L) = A + BL(L+1)$ in the $g^{th}$ Landau level on a Haldane sphere defined by $Q$, with $l = Q + g$, it is simple enough to directly calculate the $V_k$ coefficients for all values of $k$, $0\leq k \leq 2l$ using \eqref{eq4.4}. But, just as in the case of the analytic inversion of the pseudopotential presented in section \ref{sec:IV}, the values of $V_k$ can also be found using the orthogonality properties of the 6J symbols. We first use the following identities from table 5 and equation (6.3.2) of reference \cite{Edmonds1996} to express the pseudopotential in terms of 6J symbols:
\begin{align}
\begin{Bmatrix}
a & b & c \\
1 & c & b
\end{Bmatrix}  
&= (-1)^{a+b+c+1} \nonumber \\
&\times  \frac{2\left[ b(b+1) + c(c+1) - a(a+1) \right]}{ \left[2b(2b+1)(2b+2)2c(2c+1)(2c+2)\right]^{1/2}}  \nonumber \\
\begin{Bmatrix}
j_1 & j_2 & j_3 \\
0 & j_3 & j_2
\end{Bmatrix}
&= (-1)^{j_1+j_2+j_3}\left[ (2j_2+1) (2j_3+1)\right]^{-1/2}. \label{eq5.2}
\end{align}
With equations \eqref{eq5.2}, the harmonic pseudopotential can be expressed as a sum over only two 6J symbols,
\begin{align}
(-1)^L V_{Q,l}(L) &= (-1)^L\left[ A + BL(L+1)\right] \nonumber \\
&= c_0 
\begin{Bmatrix}
l & l & 0 \\
l & l & L
\end{Bmatrix} + c_1
\begin{Bmatrix}
l & l & 1 \\
l & l & L
\end{Bmatrix}, \nonumber \\
c_0 &= \left[ A + B 2l(l+1)\right](-1)^{2l} (2l+1), \nonumber  \\
c_1 &= B (-1)^{2l} l (2l+1) (2l+2). \label{eq5.3}
\end{align}
Substituting \eqref{eq5.3} into equation \eqref{eq4.4} and applying the orthogonality property of equation \eqref{eq4.1} gives the full set of $V_k$ coefficients,
\begin{align}
V_k =& \frac{\sqrt{Q}}{(2l+1)^2}(-1)^{2Q}
\begin{pmatrix}
 l & k & l \\
 -Q & 0 & Q
 \end{pmatrix}^{-2} \nonumber  \\
  &\times \left[ c_0 \delta_{k,0} + c_1 \delta_{k,1} \right],
\end{align}
where $\delta_{k,i}$ is the standard Kronecker delta. In other words, any harmonic pseudopotential $V_{Q,l}(L) = A + BL(L+1)$ can be written in spatial coordinates in terms of the first two Legendre polynomials alone, or equivalently, $V(r_{ij}) =  a + b \mathbf{r}_i\cdot \mathbf{r}_j$, where $a$ and $b$ are constants. This result has been found on the surface of the Haldane sphere, but mapping this result to the infinite plane does not change the form of $V(r_{12})$ because the mapping is stereographic (angle-preserving) and, at most, results in different scaling constants, $a$ and $b$. Solving the $N$-particle planar system in the presence of a harmonic pseudopotential offers an alternative confirmation of the harmonic pseudopotential theorem of W\'ojs and Quinn \cite{Wojs1998,Quinn2000} and links their works to Johnson's work on the spatial harmonic potential \cite{Johnson1991, Johnson2002}.

For the planar quantum Hall system, $N$ identical electrons are confined to the two-dimensional $x$-$y$ plane in the presence of a perpendicular magnetic field,  $\mathbf{B} = B\hat{z}$. In the symmetric gauge,  $\mathbf{A}_i = \frac{1}{2}( \mathbf{B} \times \mathbf{r}_i )$ is the vector potential for the $i^{th}$ electron, and the single-electron Hamiltonian is 
\begin{equation}
\hat{H}_{0,i} =  \frac{ 1 }{2m} \mathbf{p}^2_i+  \frac{1}{2} m \Omega_c^2  \mathbf{r}^2_i + \Omega_c  \hat{L}_{z,i} , \label{eq5.6}
\end{equation}
where $\Omega_c$ is one-half the cyclotron frequency, $\omega_c = eB/mc$. The single-particle solutions of the Schr$\ddot{o}$dinger equation, $\psi_{ n, m}(\mathbf{r})$, are quantized into Landau levels defined by the quantum numbers $n$ and $m$, with $n \geq 0$ and energies given by
\begin{equation}
E_{n, m} =  \hbar \Omega_c (2n + 1 + m + |m|). \label{eq5.7}
\end{equation}
The $g^{th}$ Landau level, LL$g$, is defined by all single particle states with the same energy, $E_g = \hbar \Omega_c(2g+1)$. Each LL is highly degenerate in states with $n = g$ and $m \leq 0$, but for $g > 0$, the LL also contains states with $0 \leq n < g$ and $0 < m \leq g$. For example, because LL1 contains all single-particle states with energy $E=\hbar\Omega_c$, it includes infinitely many states with $n=1$ and $m\leq0$ in addition to one state with $n=0$ and $m = 1$. While it would be convenient to consider only negative-$m$ states, the positive-$m$ states must also be included in our analysis for all but the lowest LL.

The corresponding N-particle problem is not analytically solvable for a general interaction potential, but it is in the case of a harmonic spatial potential \cite{Johnson1991,Johnson2002}. We have already shown that the harmonic pseudopotential $V_H(L_{ij}) = A+BL_{ij}(L_{ij}+1)$ is equal to a spatial potential $V_H(r_{ij}) = a + b \mathbf{r}_i \cdot \mathbf{r}_j$ in the planar system. The resulting problem can be more simply solved in a reference frame rotating at half of the cyclotron frequency. The Hamiltonian in this frame is
\begin{align}
\hat{H} = & \sum_{i=1}^N \frac{p_{i}^2}{2m} + \left[ \frac{1}{2}m\Omega_c^2 - \frac{b}{2} \right] 
\sum_{i=1}^N r_i^2 \nonumber \\
& + \frac{a}{2}N(N-1) + \frac{b}{2}\sum_{\text{all }i,j}^N \mathbf{r}_i \cdot \mathbf{r}_j. \label{eq5.8}
\end{align}
We can rewrite \eqref{eq5.8} in terms of one center of mass coordinate, $\xi_1$, and $N-1$ relative coordinates, $\xi_2, \xi_3, \ldots, \xi_N$. If we chose orthonormal coordinates such that
\begin{align}
\sum_{i=1}^N r_i^2 = \xi_{1}^2 + \sum_{\alpha = 2}^N \xi_{\alpha}^2 \quad  \text{ and }\quad
\xi_1 = \frac{1}{\sqrt{N}} \sum_{i = 1}^N \mathbf{r}_i ,  \label{eq5.9}
\end{align}
then \eqref{eq5.8} separates as
\begin{align}
\hat{H}_{cm} &= \frac{p_1^2}{2m} +  \frac{1}{2} m \Omega_1^2 \xi_1^2 + \frac{a}{2}N(N-1), \nonumber \\
\hat{H}_{rel} &= \sum_{\alpha = 2}^N \frac{p_\alpha^2}{2m} + 
 \frac{1}{2} m \Omega^2  \sum_{\alpha = 2}^N \xi_\alpha^2 , \label{eq5.10}
\end{align}
where we have chosen $\frac{1}{2} m\Omega_1^2 = \frac{1}{2}m\Omega_c^2  + \frac{1}{2}(N-1)b   $ and $\frac{1}{2} m\Omega^2 = \frac{1}{2}m\Omega_c^2  - \frac{1}{2}b   $. Note that, while the choice of the center of mass coordinate is unique, there are many possible choices for defining the relative coordinates. Regardless of the choice of relative coordinates, all relative normal modes share the same oscillator frequency, $\Omega$, as is the case in other harmonic systems \cite{Zaluska-Kotur2000}. The energies of the center of mass and relative coordinates in this rotating frame are 
\begin{align}
E_{cm, rotating} &=  \hbar \Omega_1 (2n_1 + 1 + |m_1|) + \frac{a}{2}N(N-1), \nonumber \\
E_{rel, rotating} &=  \hbar \Omega \sum_{\alpha = 2}^N (2n_\alpha + 1 + |m_\alpha|).  \label{eq5.11}
\end{align}
Rotating back to the laboratory frame introduces a factor of $\frac{1}{2} \hbar \Omega_c m_\alpha$ for each normal mode,
\begin{align}
E_{cm, lab} =&   \hbar \Omega_1 (2n_1 + 1 + |m_1| + m_1)  \nonumber \\
		& + \frac{a}{2}N(N-1)   + \hbar (\Omega_c - \Omega_1)m_1, \nonumber \\
E_{rel, lab} = &\hbar \Omega \sum_{\alpha = 2}^N (2n_\alpha + 1 + |m_\alpha| + m_\alpha) \nonumber \\
		&+\hbar (\Omega_c - \Omega)\sum_{\alpha = 2}^N m_\alpha. \label{eq5.12}
\end{align}
While the total azimuthal angular momentum of the system, $M_{total} = \sum_{i = 1}^N m_i$ is unchanged by the transformation to normal modes, the Landau level occupancy of the normal modes need not match the LL occupancy in the independent particle picture. For example, if we take $N$ particles all in the $g^{th}$ LL in the independent particle picture and transform to normal modes, the normal modes may not all lie within LL $g$, even in the absence of interactions.

Now, consider two different states in the absence of interactions, $\Psi$ and $\Psi'$, each with $N$ electrons all in the same Landau level in the independent particle picture. Assume they both have the same center of mass energies in the non-interacting system with $m_{cm} = m_{cm}'$. Because all $N$ particles lie in the same Landau level, the total initial energies are the same, $E_0 = E_0'$. Since $E = E_{cm} + E_{rel}$, the initial relative energies are also the same, and in terms of the normal mode quantum numbers,
\begin{align}
(2n_1 + 1 + m_1 + |m_1|) &= (2n_1' + 1 + m_1' + |m_1'|),  \nonumber \\
\sum_{\alpha = 2}^N (2n_\alpha + 1 + m_\alpha + |m_\alpha|) &= \sum_{\alpha = 2}^N(2n_\alpha' + 1 + m_\alpha' + |m_\alpha'|). \label{eq5.13}
\end{align}
If we then turn on a harmonic interaction potential, the difference in the energies of $\Psi$ and $\Psi'$ is 
\begin{equation}
E - E' =  \hbar (\Omega - \Omega_c) (M_{rel} - M_{rel}'),  \label{eq5.14}
\end{equation}
where $M_{rel} $ is the sum of the azimuthal quantum numbers, $m_\alpha$, of the relative coordinates. According to \eqref{eq5.14}, a harmonic pseudopotential will separate the energies of the initial states with different relative azimuthal angular momenta, $M_{rel}$, but will not break the degeneracy of states with the same $M_{rel}$. 

Mapping this result back to the sphere for comparison to the harmonic pseudopotential theorem is straightforward because the relative, center of mass, and total azimuthal quantum numbers, $M_{rel}, M_{cm}$, and $M_{total}$, respectively, connect to the total $N$-particle angular momentum $L$, the total azimuthal angular momentum $L_z$, and the shell angular momentum $l$ on the sphere by way of the following relations \cite{Fano1986, Quinn2009}:
\begin{align}
M_{total} = Nl+L_z, \quad M_{rel} = Nl - L, \quad M_{cm} = L + L_z.
\end{align}
Since $l$ is a constant for $N$ particles in the same Landau level, the result in \eqref{eq5.11} when mapped back to the sphere indicates that a harmonic pseudopotential does not break the degeneracy of $N$-particle states with the same total angular momentum, $L$. The difference between the planar system and the spherical system is that states with different $M_{cm}$ and equivalent $M_{rel}$ will have different energies on the plane, whereas the azimuthal quantum number $L_z$ does not affect the energy on the sphere.

The harmonic pseudopotential is of particular interest because it is a dividing line between two major categories of electron correlations, Laughlin and pairing. A pseudopotential that increases more rapidly than a harmonic pseudopotential for the highest allowed pair angular momenta will cause the electrons to avoid forming pair states with the highest angular momentum. In such a system, the lowest energy states, then, will be Laughlin correlated. In contrast, if the pseudopotential increases less rapidly than a harmonic pseudopotential for high pair angular momenta, electrons in the system will favour pairing, an the lowest energy states will be pair correlated.

\section{\label{sec:VI}Examples}
In general, any pseudopotential for a single finite Landau level can be inverted to give a corresponding effective spatial potential according to the procedure of section \ref{sec:IV}. However, in some cases, calculation of the effective spatial potential coefficients can lead to spatial potentials that are difficult to interpret. Sometimes, the calculated $V_k$ coefficients will be dramatically large for the larger values of $k$. The resulting effective potential from equation \eqref{eq2.11} will exhibit strong ringing, and such extreme oscillations in the plot of the potential as a function of $\theta_{12}$ or $r_{12}$ are naturally difficult to interpret as a realistic model spatial potential.

Fortunately, this ringing behavior is often a result of one of several factors that can be addressed directly. We will discuss these factors in the following subsections using examples that show how to interpret the effective potential when such strong ringing appears in an inversion calculation.

\subsection{\label{sec:vi.i}Numerical Error}
The first and most important possible source of strong spatial potential oscillations is numerical inaccuracy. As mentioned in section \ref{sec:IV}, it is straightforward to analytically invert a Coulomb pseudopotential. If the pseudopotential terms for a given LL are evaluated exactly without numerical approximations according to equation \eqref{eq3.9}, then substituting these values into equation \eqref{eq4.4} gives the $V_k$ coefficients as large analytic sums over exact terms. The analytic sums can be evaluated algebraically without numerical approximations, and the resulting $V_k$ coefficients will simplify to $V_k = 1$ for all $0 \leq k \leq 2l$, as expected.

However, if the terms of the Coulomb pseudopotential are first calculated numerically, the sums of equation \eqref{eq4.4} may diverge, sometimes wildly, from the expected results due to rounding errors. Unlike the sums of equation \eqref{eq3.9}, the sums in \eqref{eq4.4} often feature the addition and subtraction of very large numbers, the majority of which nearly cancel one another. If the computer does not store enough digits of the approximate terms in equation \eqref{eq4.4}, then the computer's summation will be dominated by numerical rounding error. This particular error appears most dramatically for the coefficients with larger $k$, for which the 3J symbols in the denominators of equation \eqref{eq4.4} are sometimes very small. For example, in a typical machine-precision, floating-point calculation using the pseudopotential in the lowest LL for Q = 30, equation \eqref{eq4.4} gives $V_k \approx 1$ to within 6 significant figures for the first 38 terms. The following several terms are also near $1$, but then diverge wildly from $1$ dramatically until the final term, $V_{60} \approx -1.80599\times10^{18}$. A plot of an effective potential using these coefficients will be dramatically oscillatory and obviously does not accurately reflect the Coulomb potential.

This issue can be very simply dealt with by simply performing calculations using sufficient numerical accuracy. In the Q = 30 system, for example, the correct result is recovered by enforcing an appropriate level of numerical accuracy. Numerical inaccuracies of this type can also be reduced in some cases by combining and simplifying large sums of fractions arithmetically before evaluating the expressions numerically, although this method is not always sufficient or helpful.

\subsection{\label{sec:vi.ii}Discontinuous potentials}
A second source of potential ringing occurs when the pseudopotential represents a discontinuous or cusped potential. While the partial wave expansion can be used to describe any number of smooth, continuous spatial potentials, functions with discontinuities or sharp cusps are less well represented by a sum over Legendre polynomials, and their expansions typically show very strong ringing. The canonical example of such a discontinuous function would be the hard sphere or delta function pseudopotential, $V_{\delta}(L) = \delta_{2l-L,1}$, for which the Laughlin wave function is famously a solution\cite{Haldane1983}. Inverting this pseudopotential in the lowest LL yields a set of $V_k$ coefficients that become increasingly negative with increasing $k$ starting with the very first term, as shown in figure \ref{fig2} for the $Q = 20$ system. The resulting effective potential is extremely oscillatory, but in this case the extreme ringing is the most accurate possible representation of the effective potential in terms of partial waves. The extreme oscillations arise because the actual effective potential is too sharp to approximate well with a non-infinite partial wave expansion. 

\begin{figure}
                \centering
                \includegraphics[width=.5\textwidth]{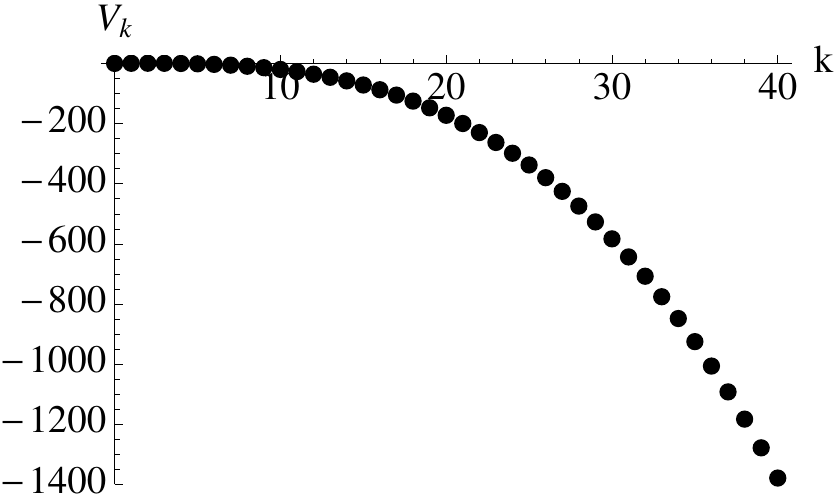}
                \label{fig2}
\caption{The $V_k$ coefficients for $Q=20$ corresponding to a delta pseudopotential, $V_{\delta}(L) = \delta_{2l-L,1}$ in the lowest Landau level. A plot of the resulting $V_{eff}(\theta)$ is too extremely oscillatory to be informative. The extreme ringing here arises because the spatial potential that produces a delta pseudopotential is too sharp to be modeled well by a partial wave expansion.
}
\label{fig2}
\end{figure}

While it is difficult to describe the resulting effective potential exactly based on the appearance of the plot, the potential behavior can be examined indirectly by numerical integration of the effective potential over small angles. The potential is very strong in the very small $\theta_{12}$ regime and weak for larger $\theta{12}$ in relatively small finite systems, and this behavior becomes more dramatic with increasing system size. In other words, a true delta function pseudopotential likely arises from a pure delta spatial potential, $V(\theta_{12}) \propto \delta(\theta_{12})$, and the extreme oscillations of the spatial potential arise because finite partial wave expansions are ill suited to express discontinuous functions such as a hard shell spatial potential.

\subsection{\label{sec:vi.ii}Sensitivity of high-$V_k$ terms}

Even if numerical accuracy has been addressed and the pseudopotential is not the result of a discontinuity, the highest $k$ terms of the inversion are very sensitive to minor variations in the pseudopotential, and this sensitivity is a third cause of strong ringing. Ringing of this type can be reduced by either restricting the finite system size or by simply ignoring the divergent high-$k$ $V_k$ coefficients.

For example, consider a Gaussian pseudopotential in the lowest Landau level, $V_{Gauss}(L) = 0.9\times 2^{-0.5 (2Q-L)^2}$, which is on roughly the same scale as the Coulomb pseudopotential. Inverting this pseudopotential for a $Q=30$ system in LL0 gives $V_k$ values that appear to asymptotically approach zero until around $k = 45$ after which they diverge from zero. As before, an effective pseudopotential with large coefficients for large $k$ is highly oscillatory, but a great deal of these terms can be avoided by simply reducing the system size. Inverting the same pseudopotential for a $Q=20$ system gives the same behavior as the $Q=30$ but lacks any highly divergent larger $k$ terms. By considering the $Q=20$ system instead of the larger $Q=30$ system, it can be seen that a Gaussian pseudopotential is produced by a softened repulsive step function, as shown in figure \ref{fig3}.

An even simpler way to eliminate ringing to examine the form of the pseudopotential is to simply ignore the divergent high-$k$ terms of the effective potential. For example, when inverted in a $Q=30$ LL0 system, the exponential pseudopotential $V_{exp}(L) = 0.15 + e^{(-2Q+L)}$ $V_k$ coefficients are well-behaved and asymptotically approach 0 for $0 \leq k \leq 54$, but diverge for higher values of $k$ (e.g. $V_{59} = 3.20112\times 10^{5}$). This spatial potential is, as expected, highly oscillatory, but if we simply ignore the last 6 $V_k$ coefficients in the effective potential sum, the resulting plot is very well-behaved (see figure \ref{fig4}. We can verify that these high-$k$ terms are not contributing strongly to the actual pseudopotential behavior by computing the pseudopotential twice with equation \eqref{eq3.9}, once using all calculated $V_k$ coefficients, and once using the same $V_k$ coefficients with the final six divergent coefficients set to zero. The resulting pseudopotentials are not numerically identical, but their differences are extremely small, on the order of $10^{-28}$ or less for all values. The high $V_k$ terms are numerically highly sensitive to very minor changes to the pseudopotential, but contribute little to the pseudopotential, and can be safely ignored when the general asymptotic behavior is clearly demonstrated by earlier $V_k$ terms.  

\begin{figure}
                \centering
                \includegraphics[width=.5\textwidth]{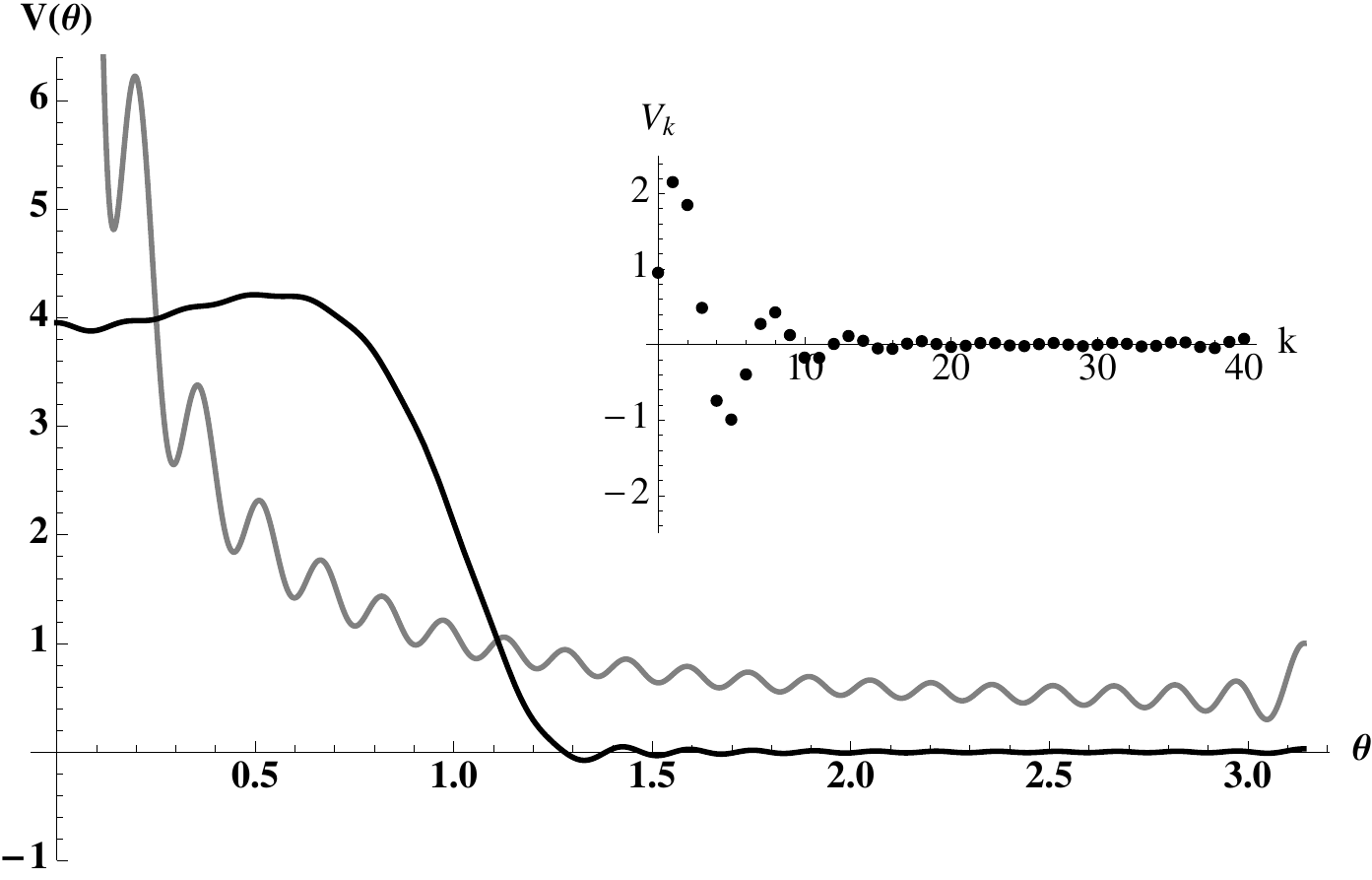}
                \label{fig3}
\caption{The $V_k$ coefficients (inset) and the resulting effective spatial potential (black curve), $V_{eff}(\theta)$ for the Gaussian pseudopotential $V_{Gauss}(L) = 0.9\times 2^{-0.5 (2Q-L)^2}$ in the lowest Landau level for $Q = 20$. The Coulomb potential for the $Q=20$ system is shown in grey for comparison.
}
\label{fig3}
\end{figure}

\begin{figure}
                \centering
                \includegraphics[width=.5\textwidth]{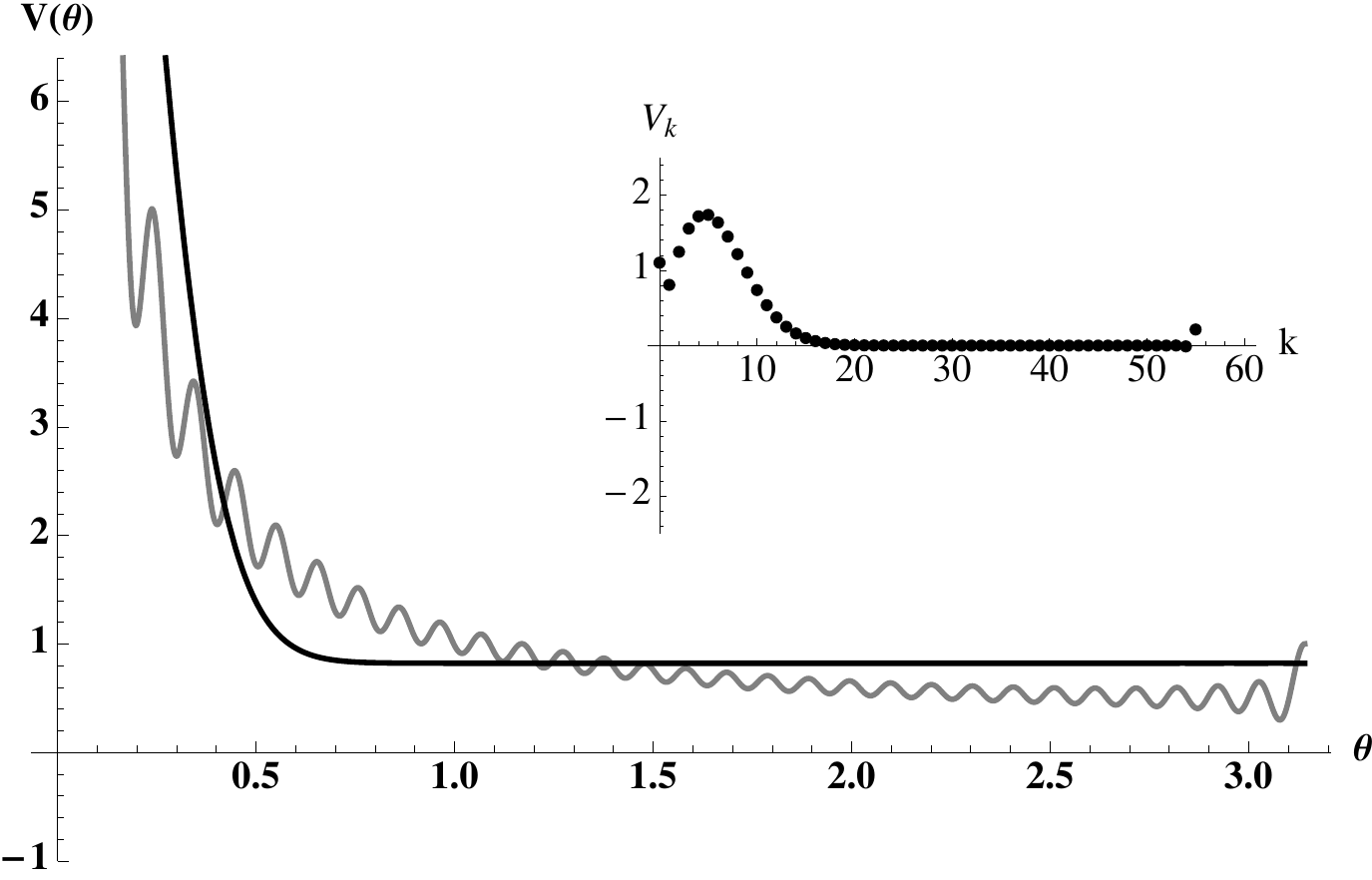}
                \label{fig4}
\caption{The $V_k$ coefficients (inset) and the resulting effecitve spatial potential (black curve), $V_{eff}(\theta)$ for the exponential pseudopotential $V_{exp}(L) = 0.15 + e^{(-2Q+L)}$ in the lowest Landau level for $Q=30$. The $V_k$ values for $k \geq 55$ are too divergent to be shown to scale and are set to zero to give the effective potential plot shown here. These divergent high-$k$ coefficients are overly sensitive to minor numerical changes to the pseudopotential and setting them equal to zero changes the pseudopotentials very little. The Coulomb potential for the $Q = 30$ system is shown in grey for comparison. 
}
\label{fig4}
\end{figure}

\section{\label{sec:VII}Conclusions}

The Haldane spherical geometry is particularly useful because it allows us to take full advantage of the well-understood theories of angular momentum algebra, as we have shown in the derivations here. Describing the spatial potential in terms of a partial wave expansion enables the modeling of a general spatial potential rather than only the Coulomb potential. While obviously the Coulomb potential describes the actual interactions between particles in the quantum Hall system, additional system properties can be incorporated into a modified Coulomb potential as an approximation to the actual experimental conditions. In particular, as indicated in section \ref{sec:II}, the experimental system is only nearly two-dimensional, and the effect of the third-dimension confinement can be incorporated into the perfectly two-dimensional model as a modification to the Coulomb potential through the $V_k$ coefficients. The connection between the spatial potential and the physical confinement suggests a way to tune the system behaviors in the quantum Hall effect, and it may be possible to select specific third-dimensional confinements to produce the set of $V_k$ parameters that would alter the system behavior in desired ways.

In addition, by connecting the spatial potential to the pseudopotential, we also are able to describe any desired model pseudopotential in terms of the spatial potential and the $V_k$ parameters. This technique could be useful for producing appropriate model spatial potentials used in quantum Monte Carlo simulations of the quantum Hall effect. In addition, connecting the spatial potential and pseudopotential allows us to show why the harmonic pseudopotential fails to break angular momentum degeneracy in the quantum Hall system, in confirmation of the harmonic pseudopotential theorem. It is also interesting to note that, in the case of non-harmonic pseudopotentials, a specific pseudopotential will be produced by different spatial functions in different Landau levels.

While the pseudopotential is generally given in terms of the pair interactions, it is also possible to describe the three-body interactions in a similar way with the three-body pseudopotential. For example, the famous Moore-Read wave function is a many-body solution to a three-body, rather than a two-body, pseudopotential. Ideally, we would like to be able to invert the three-body pseudopotentials in order to obtain the two-body pseudopotentials through a similar analytic process as presented in section \ref{sec:IV}, but the situation is rather more complicated and requires the use of the coefficients of fractional parentage. Additionally, not all hypothetical three-body pseudopotentials can be constructed from a two-body pseudopotential, so a three-body pseudopotential inversion is likely to be of less use.

\begin{acknowledgments}
We gratefully acknowledge support by the Office of Basic Energy Sciences, U.S. Department of Energy, through a grant (DE-FG02-02ER15283) to the University of Tennessee. In addition, we also thank Dr. John Quinn for various discussions on the topic.

\end{acknowledgments}

\appendix

\section{\label{appendix:i}Details of the two-body matrix element derivation}
Evaluating the matrix element of equation \eqref{eq2.5} for each value of $k$ requires the separation of variables for particles $1$ and $2$, which can be done using the spherical harmonic addition theorem, equation (4.6.6) in \cite{Edmonds1996},
\begin{equation}
P_k(\cos \theta_{12}) = \frac{4\pi}{2k+1}\sum_{m = -k}^{k} Y^*_{k,m}(\Omega_1) Y_{k,m}(\Omega_2).\label{eq2.6}
\end{equation}
$\Omega_i$ is an abbreviation for the set of angular variables $\theta_i, \phi_i$ for the $i^{th}$ particle. Incorporating equation \eqref{eq2.6} into equation \eqref{eq2.5} yields the following integral equation in Coulomb units
\begin{align}
\langle  l_1', m_1' ; &  l_2', m_2' | V(r_{12})|  l_1,m_1 ;  l_2, m_2 \rangle =  \nonumber \\
&\frac{1}{ \sqrt{Q}} \sum_{k = 0}^{\infty} \sum_{m = -k}^k  V_k \left( \frac{4\pi}{2k+1} \right) \nonumber \\
&\times \int Y^*_{Q,l_1',m_1'}(\Omega_1) Y^*_{k,m}(\Omega_1) Y_{Q,l_1,m_1}(\Omega_1) d\Omega_1 \nonumber \\
& \times \int Y^*_{Q,l_2',m_2'}(\Omega_2) Y_{k,m}(\Omega_2) Y_{Q,l_2,m_2}(\Omega_2) d\Omega_2		\label{eq2.7}
\end{align}
The monopole quantum number, $Q$, has been dropped from the state-vector notation for brevity. The integrals in equation \eqref{eq2.7} can be evaluated using the results from reference \cite{Wu1977}. First, we rewrite the spherical harmonics in terms of monopole harmonics using the simple relation $Y_{k,m} = Y_{0,k,m}$. We also use \textit{theorem 1} from \cite{Wu1977},   
\begin{equation}
Y^*_{Q,l,m} = (-1)^{Q+m}Y_{-Q, l, -m}, \label{eq2.8}
\end{equation}
to rewrite the complex conjugate monopole harmonics as non-conjugate terms. 

Evaluating the modified version of \eqref{eq2.7} then requires a closer examination of \textit{Theorem 3} from \cite{Wu1977}. Although \textit{Theorem 3} requires $m+m'+m'' = 0$ as a necessary condition for evaluating the integrals, the subsequent proof actually does not rely on the assumption, although it does require $q+q'+q''=0$. When $q+q'+q'' = 0$, the integral is non-zero only when $m+m'+m'' = 0$, but this is a consequence of monopole harmonics' relation to the standard rotation functions. In one region of the sphere, for example, Wu and Yang define the monopole harmonics in terms of the rotation functions of reference \cite{Edmonds1996}, 
\begin{equation}
Y_{Q,l,m} = \left[ \frac{2l+1}{4\pi} \right] e^{i (q+m)\phi} d_{-m,q}^{(l)}{\theta}. \label{eqA4}
\end{equation}
The integral over the $\phi$ variable in \textit{Theorem 3} gives $0$ unless $m+m'+m''+q+q'+q'' = 0$. Since $q+q'+q'' = 0$ is already required, it follows that the integral is zero except when the sum over the $m$'s is zero. This information is already contained in the 3J symbols, which are zero under the same circumstances.

Combining the theorems of Wu and Yang with \eqref{eq2.7} yields
\begin{align}
\langle  &l_1', m_1';   l_2', m_2' | V(r_{12})|  l_1, m_1;   l_2, m_2 \rangle = \nonumber \\
&\frac{\bar{l}(-1)^{\eta } }{\sqrt{Q}} \sum_{k = 0}^{\infty}\sum_{m = -k}^{k} 
\left[ V_k 
\begin{pmatrix}
l_1' & k & l_1 \\
-m_1' & -m & m_1
\end{pmatrix} \right. \nonumber \\
&\times \left.
\begin{pmatrix}
l_1' & k & l_1 \\
-Q & 0 & Q
\end{pmatrix} 
\begin{pmatrix}
l_2' & k & l_2\\
-m_2' & m & m_2
\end{pmatrix}
\begin{pmatrix}
l_2' & k & l_2 \\
-Q & 0 & Q
\end{pmatrix}\right], \label{eq2.9}
\end{align}
where $\eta = 2Q + m_1' + m + m2' + l_1'+ l_2'+ l_1 + l_2 $ and $\bar{l} = \left[(2l_1'+1)(2l_1+1)(2l_2'+1)(2l_2+1)\right]^{\frac{1}{2}}$. The sums over $m$ and $k$ simplify according to the properties of the 3J symbols. Since the 3J symbols are zero unless the sum over the lower three indices is zero, the sum over $m$ collapses with $m = m_1-m_1' = m_2'-m_2$. In addition, the 3J symbols are non-zero only when the upper three indices satisfy the triangle condition, $|l_1' - l_1| \leq k \leq l_1+l_1'$. The triangle inequality requirement reduces the infinite sum over $k$ to a finite sum up to $k_{max}$, where $k_{max}$ is the largest integer that satisfies both $k_{max} \leq l_1+l_1'$ and $k_{max} \leq l_2+l_2'$. Applying all these reductions to \eqref{eq2.9}, we get the final expression for a two-particle matrix element on the Haldane sphere:
\begin{widetext}
\begin{align}
\langle  l_1', m_1';  l_2', m_2' | V(r_{12})| l_1, m_1;  l_2, m_2 \rangle = 
\frac{\bar{l} (-1)^{\eta} }{ \sqrt{Q}}\sum_{k = 0}^{k_{max}}  V_k   
\begin{pmatrix}
l_1' & k & l_1 \\
-m_1' & -m & m_1
\end{pmatrix}
\begin{pmatrix}
l_1' & k & l_1 \\
-Q & 0 & Q
\end{pmatrix}     
\begin{pmatrix}
l_2' & k & l_2\\
-m_2' & m & m_2
\end{pmatrix}
\begin{pmatrix}
l_2' & k & l_2 \\
-Q & 0 & Q
\end{pmatrix}.     \label{eqA7}
\end{align}  
\end{widetext}
The value of $\bar{l}$ is still given by $\bar{l} = \left[(2l_1'+1)(2l_1+1)(2l_2'+1)(2l_2+1)\right]^{\frac{1}{2}}$, but since $m$ is now limited by $m = m_1-m_1' = m_2'-m_2$, the value of $\eta$ is now given by $\eta = 2Q + m2' + m_1 + l_1'+ l_2'+ l_1 + l_2 $. The matrix element is given in units of the Coulomb energy, $e^2/4\pi\epsilon\lambda_0$, where $\lambda_0 = \sqrt{\hbar c/eB}$.

\section{\label{appendix:ii}Symmetric and antisymmetric pair states}
In this discussion, the pair angular momentum eigenstates, $| Q, l_1, l_2; L ,M\rangle$, have been neither symmetrized nor antisymmetrized so that equations \eqref{eq3.8}, and \eqref{eq3.9} could be applied to fermion, boson, or distinguishable particle pairs. Construction of the appropriate antisymmetric and symmetric eigenstates requires knowledge of the coupled monopole harmonics given in appendix D of \cite{Wu1976}. This wave function, $F_{Q,Q',L,M}(\theta_1,\phi_1,\theta_2,\phi_2)$, is the projection into coordinate space of the ket $| Q, l_1, l_2; L ,M\rangle$. The coupled monopole harmonics behave under rotations similarly to the coupled spherical harmonics and obey the following equation:
\begin{align}
F_{Q,Q',L,M}(\theta_1,\phi_1, & \theta_2,\phi_2) = \nonumber \\
\sum_{m,m'} &Y_{Q,l,m}(\theta,\phi) Y_{Q',l',m'}(\theta', \phi')  \nonumber \\ 
& \times \langle l,l',L,M | l,m,l',m' \rangle, \label{eqB1}
\end{align}
where $\langle l,l',L,M | l,m,l',m' \rangle$ is an ordinary Clebsch-Gordan coefficient. Antisymmetric and symmetric states can be derived using the symmetry properties of the Clebsch-Gordan coefficients. Applying the two-particle permutation operator, $\hat{P}_{1,2}$, to \eqref{eqB1} gives
\begin{align}
\hat{P}_{1,2} F_{Q,Q',L,M}(\theta_1,\phi_1,\theta_2,\phi_2)  =  (-1)^{l+l'-L}  \nonumber \\
\times \sum_{m,m'} \left[ Y_{Q,l,m}(\theta_2,\phi_2) Y_{Q',l',m'}(\theta_1,\phi_1) \right. \nonumber \\
 \left. \times \langle l',l,L,M | l',m',l,m \rangle \right], \label{eqB3}
\end{align}
or in terms of kets,
\begin{equation}
\hat{P}_{1,2}  | l, l'; L, M \rangle = (-1)^{l+l'-L} | l', l; L, M \rangle.  \label{eqB4}
\end{equation}
The antisymmetric/symmetric pair eigenstates can be formed by operating on the original basis state, $| l, l'; L, M\rangle$ with the antisymmetrization/symmetrization operators, $\hat{A}/\hat{S}$, where
\begin{equation}
\hat{A} = \mathcal{N}(1- \hat{P}_{1,2}), \quad \hat{S} = \mathcal{N} (1 + \hat{P}_{1,2}). \label{eqB5}
\end{equation}
Here, $\mathcal{N}$ is an appropriate normalization constant, which will depend on whether $l = l'$. A more complete discussion of the symmetry of pair and multi-particle states can be found in \cite{Fano1959}.

\section{\label{appendix:iii}Details of the pair pseudopotential derivation}

In order to evaluate equation \eqref{eq3.2}, it is convenient to express the problem in terms of tensor operators and the Racah functions, $C_{m}^{(k)}$. The spherical harmonic addition theorem of \eqref{eq2.6} can be expressed as a scalar product of two rank $k$ tensor operators, $\mathbf{C}^{(k)}$, of different arguments,  
\begin{equation}
P_k(\cos \theta_{12}) = \left[ \mathbf{C}^{(k)}(\theta_1, \phi_1) \cdot  \mathbf{C}^{(k)}(\theta_2, \phi_2)  \right]_0^{(0)}. \label{eq3.3}
\end{equation}
Substituting \eqref{eq3.3} into \eqref{eq3.2}, the matrix element of the right side of \eqref{eq3.2} for a single value of $k$ is given by Equation (7.1.6) of reference \cite{Edmonds1996}:
\begin{align}
\langle l_1', l_2'; L', M' | & \left[ \mathbf{C}^{(k)}(\Omega_1) \cdot  \mathbf{C}^{(k)}(\Omega_2)  \right]_0^{(0)}  |  l_1, l_2; L, M \rangle = \nonumber \\
(-1)^{l_1 +l_2' + L} & \delta_{L, L'}  \delta_{M,M'} 
\begin{Bmatrix}
L & l_2' & l_1' \\
k & l_1 & l_2
\end{Bmatrix} \nonumber \\
\sum_\gamma& (Q, l_1' || \mathbf{C}^{(k)}(\Omega_1) || \gamma,  l_1 ) \nonumber \\
& \times (\gamma, l_2' || \mathbf{C}^{(k)}(\Omega_2) || Q, l_2). \label{eq3.4}
\end{align}
The quantity in curly braces is a Wigner 6J symbol. The expression $(Q, l_1' || \mathbf{C}^{(k)}(\Omega_1)|| \gamma,  l_1 )$ is a reduced matrix element which is related to an ordinary matrix element by the Wigner-Eckart theorem, given by equation (5.4.1) in \cite{Edmonds1996} as
\begin{align}
\langle Q, l', m' | C_x^{(k)} | \gamma, l, m \rangle = &(-1)^{l'-m'} 
\begin{pmatrix}
l' & k & l \\
-m' & x & m
\end{pmatrix} \nonumber \\
& \times ( Q, l' || \mathbf{C}^{(k)} || \gamma, l ). \label{eq3.5}
\end{align}
The expression on the left of \eqref{eq3.5} is, by definition, given by the integral equation
\begin{align}
\langle Q, l', m' | C_x^{(k)} | \gamma, l, m \rangle  &= \nonumber \\
\left( \frac{4\pi}{(2k+1)} \right)^{\frac{1}{2}} & \int Y^*_{Q,l',m'} Y_{0,k,x} Y_{\gamma,l,m} d\Omega,	 \label{eq3.6}
\end{align}
which can be evaluated using \textit{theorem 3} of \cite{Wu1977}. Combining equations \eqref{eq3.5} and \eqref{eq3.6} with the results of Wu and Yang gives the following expression for the reduced matrix element:
\begin{align}
( Q, l' || \mathbf{C}^{(k)} || & \gamma, l ) = (-1)^{Q+2m'+l+k} \nonumber \\
& \times \left[ (2l'+1)(2l+1) \right]^{\frac{1}{2}} 
\begin{pmatrix}
l' & k & l \\
-Q & 0 & \gamma
\end{pmatrix}.  \label{eq3.7}
\end{align}

By the symmetry properties of the 3j symbols, the reduced matrix element in \eqref{eq3.7} is only non-zero when $\gamma = Q$. As a result, using \eqref{eq3.7} in \eqref{eq3.4} collapses the sum over $\gamma$. Incorporating the simplified \eqref{eq3.4} into \eqref{eq3.2} and including the triangle inequality property of the 3J symbols to terminate the infinite sum over $k$ gives the matrix element in the coupled angular momentum basis as a finite sum for particles in any Landau level,
\begin{widetext}
\begin{align}
 \langle Q, l_1', l_2'; L', M' | V(r_{12}) | Q, l_1, l_2; L, M \rangle = 
\delta_{L, L'}  \delta_{M,M'} \frac{\bar{l}(-1)^{\eta}}{\sqrt{Q}}\sum_{k=0}^{k_{max}} V_k   
\begin{Bmatrix}
 L & l_2' & l_1' \\
 k & l_1 & l_2
 \end{Bmatrix}
 \begin{pmatrix}
 l_1' & k & l_1 \\
 -Q & 0 & Q
 \end{pmatrix}
 \begin{pmatrix}
 l_2' & k & l_2 \\
 -Q & 0 & Q
 \end{pmatrix}. \label{eqB6}
\end{align}
\end{widetext}
Here, $\eta = 2Q+L+2l_1+l_2+l_2'$, and $\bar{l} = \left[(2l_1'+1)(2l_1+1)(2l_2'+1)(2l_2+1)\right]^{\frac{1}{2}}$. The value of $k_{max}$ is determined by the triangle inequalities for the sets $\{ l_1', k, l_1 \}$ and $\{ l_2', k, l_2 \}$ as it was in section \ref{sec:II} and appendix \ref{appendix:i}.

\bibliographystyle{apsrev}
\bibliography{bibliography}

\begin{thebibliography}{33}
\expandafter\ifx\csname natexlab\endcsname\relax\def\natexlab#1{#1}\fi
\expandafter\ifx\csname bibnamefont\endcsname\relax
  \def\bibnamefont#1{#1}\fi
\expandafter\ifx\csname bibfnamefont\endcsname\relax
  \def\bibfnamefont#1{#1}\fi
\expandafter\ifx\csname citenamefont\endcsname\relax
  \def\citenamefont#1{#1}\fi
\expandafter\ifx\csname url\endcsname\relax
  \def\url#1{\texttt{#1}}\fi
\expandafter\ifx\csname urlprefix\endcsname\relax\def\urlprefix{URL }\fi
\providecommand{\bibinfo}[2]{#2}
\providecommand{\eprint}[2][]{\url{#2}}

\bibitem[{\citenamefont{Klitzing et~al.}(1980)\citenamefont{Klitzing, Dorda,
  and Pepper}}]{Klitzing1980}
\bibinfo{author}{\bibfnamefont{K.~v.} \bibnamefont{Klitzing}},
  \bibinfo{author}{\bibfnamefont{G.}~\bibnamefont{Dorda}}, \bibnamefont{and}
  \bibinfo{author}{\bibfnamefont{M.}~\bibnamefont{Pepper}},
  \bibinfo{journal}{Phys. Rev. Lett.} \textbf{\bibinfo{volume}{45}},
  \bibinfo{pages}{494} (\bibinfo{year}{1980}),
  \urlprefix\url{http://link.aps.org/doi/10.1103/PhysRevLett.45.494}.

\bibitem[{\citenamefont{MacDonald and Aers}(1984)}]{MacDonald1984}
\bibinfo{author}{\bibfnamefont{A.~H.} \bibnamefont{MacDonald}}
  \bibnamefont{and} \bibinfo{author}{\bibfnamefont{G.~C.} \bibnamefont{Aers}},
  \bibinfo{journal}{Phys. Rev. B} \textbf{\bibinfo{volume}{29}},
  \bibinfo{pages}{5976} (\bibinfo{year}{1984}),
  \urlprefix\url{http://link.aps.org/doi/10.1103/PhysRevB.29.5976}.

\bibitem[{\citenamefont{Zhang and Das~Sarma}(1986)}]{Zhang1986}
\bibinfo{author}{\bibfnamefont{F.~C.} \bibnamefont{Zhang}} \bibnamefont{and}
  \bibinfo{author}{\bibfnamefont{S.}~\bibnamefont{Das~Sarma}},
  \bibinfo{journal}{Phys. Rev. B} \textbf{\bibinfo{volume}{33}},
  \bibinfo{pages}{2903} (\bibinfo{year}{1986}),
  \urlprefix\url{http://link.aps.org/doi/10.1103/PhysRevB.33.2903}.

\bibitem[{\citenamefont{Park and Jain}(1998)}]{Park1998PRL}
\bibinfo{author}{\bibfnamefont{K.}~\bibnamefont{Park}} \bibnamefont{and}
  \bibinfo{author}{\bibfnamefont{J.~K.} \bibnamefont{Jain}},
  \bibinfo{journal}{Phys. Rev. Lett.} \textbf{\bibinfo{volume}{81}},
  \bibinfo{pages}{4200} (\bibinfo{year}{1998}),
  \urlprefix\url{http://link.aps.org/doi/10.1103/PhysRevLett.81.4200}.

\bibitem[{\citenamefont{Peterson et~al.}(2008)\citenamefont{Peterson,
  Jolicoeur, and Das~Sarma}}]{Peterson2008}
\bibinfo{author}{\bibfnamefont{M.~R.} \bibnamefont{Peterson}},
  \bibinfo{author}{\bibfnamefont{T.}~\bibnamefont{Jolicoeur}},
  \bibnamefont{and}
  \bibinfo{author}{\bibfnamefont{S.}~\bibnamefont{Das~Sarma}},
  \bibinfo{journal}{Phys. Rev. B} \textbf{\bibinfo{volume}{78}},
  \bibinfo{pages}{155308} (\bibinfo{year}{2008}),
  \urlprefix\url{http://link.aps.org/doi/10.1103/PhysRevB.78.155308}.

\bibitem[{\citenamefont{Yoshioka}(1984)}]{Yoshioka1984}
\bibinfo{author}{\bibfnamefont{D.}~\bibnamefont{Yoshioka}},
  \bibinfo{journal}{J. Phys. Soc. Jpn.} \textbf{\bibinfo{volume}{53}},
  \bibinfo{pages}{3740} (\bibinfo{year}{1984}).

\bibitem[{\citenamefont{Rezayi and Haldane}(1990)}]{Rezayi1990}
\bibinfo{author}{\bibfnamefont{E.~H.} \bibnamefont{Rezayi}} \bibnamefont{and}
  \bibinfo{author}{\bibfnamefont{F.~D.~M.} \bibnamefont{Haldane}},
  \bibinfo{journal}{Phys. Rev. B} \textbf{\bibinfo{volume}{42}},
  \bibinfo{pages}{4532} (\bibinfo{year}{1990}),
  \urlprefix\url{http://link.aps.org/doi/10.1103/PhysRevB.42.4532}.

\bibitem[{\citenamefont{W\'ojs and Quinn}(2006)}]{Wojs2006}
\bibinfo{author}{\bibfnamefont{A.}~\bibnamefont{W\'ojs}} \bibnamefont{and}
  \bibinfo{author}{\bibfnamefont{J.~J.} \bibnamefont{Quinn}},
  \bibinfo{journal}{Phys. Rev. B} \textbf{\bibinfo{volume}{74}},
  \bibinfo{pages}{235319} (\bibinfo{year}{2006}),
  \urlprefix\url{http://link.aps.org/doi/10.1103/PhysRevB.74.235319}.

\bibitem[{\citenamefont{Bishara and Nayak}(2009)}]{Bishara2009}
\bibinfo{author}{\bibfnamefont{W.}~\bibnamefont{Bishara}} \bibnamefont{and}
  \bibinfo{author}{\bibfnamefont{C.}~\bibnamefont{Nayak}},
  \bibinfo{journal}{Phys. Rev. B} \textbf{\bibinfo{volume}{80}},
  \bibinfo{pages}{121302} (\bibinfo{year}{2009}),
  \urlprefix\url{http://link.aps.org/doi/10.1103/PhysRevB.80.121302}.

\bibitem[{\citenamefont{Simon and Rezayi}(2013)}]{Simon2013}
\bibinfo{author}{\bibfnamefont{S.~H.} \bibnamefont{Simon}} \bibnamefont{and}
  \bibinfo{author}{\bibfnamefont{E.~H.} \bibnamefont{Rezayi}},
  \bibinfo{journal}{Phys. Rev. B} \textbf{\bibinfo{volume}{87}},
  \bibinfo{pages}{155426} (\bibinfo{year}{2013}),
  \urlprefix\url{http://link.aps.org/doi/10.1103/PhysRevB.87.155426}.

\bibitem[{\citenamefont{Sodemann and MacDonald}(2013)}]{Sodemann2013}
\bibinfo{author}{\bibfnamefont{I.}~\bibnamefont{Sodemann}} \bibnamefont{and}
  \bibinfo{author}{\bibfnamefont{A.~H.} \bibnamefont{MacDonald}},
  \bibinfo{journal}{Phys. Rev. B} \textbf{\bibinfo{volume}{87}},
  \bibinfo{pages}{245425} (\bibinfo{year}{2013}),
  \urlprefix\url{http://link.aps.org/doi/10.1103/PhysRevB.87.245425}.

\bibitem[{\citenamefont{Peterson and Nayak}(2013)}]{Peterson2013}
\bibinfo{author}{\bibfnamefont{M.~R.} \bibnamefont{Peterson}} \bibnamefont{and}
  \bibinfo{author}{\bibfnamefont{C.}~\bibnamefont{Nayak}},
  \bibinfo{journal}{Phys. Rev. B} \textbf{\bibinfo{volume}{87}},
  \bibinfo{pages}{245129} (\bibinfo{year}{2013}),
  \urlprefix\url{http://link.aps.org/doi/10.1103/PhysRevB.87.245129}.

\bibitem[{\citenamefont{Wooten et~al.}(2013)\citenamefont{Wooten, Macek, and
  Quinn}}]{Wooten2013}
\bibinfo{author}{\bibfnamefont{R.~E.} \bibnamefont{Wooten}},
  \bibinfo{author}{\bibfnamefont{J.~H.} \bibnamefont{Macek}}, \bibnamefont{and}
  \bibinfo{author}{\bibfnamefont{J.~J.} \bibnamefont{Quinn}},
  \bibinfo{journal}{Phys. Rev. B} \textbf{\bibinfo{volume}{88}},
  \bibinfo{pages}{155421} (\bibinfo{year}{2013}),
  \urlprefix\url{http://link.aps.org/doi/10.1103/PhysRevB.88.155421}.

\bibitem[{\citenamefont{Park et~al.}(1998)\citenamefont{Park, Melik-Alaverdian,
  Bonesteel, and Jain}}]{Park1998PRB}
\bibinfo{author}{\bibfnamefont{K.}~\bibnamefont{Park}},
  \bibinfo{author}{\bibfnamefont{V.}~\bibnamefont{Melik-Alaverdian}},
  \bibinfo{author}{\bibfnamefont{N.~E.} \bibnamefont{Bonesteel}},
  \bibnamefont{and} \bibinfo{author}{\bibfnamefont{J.~K.} \bibnamefont{Jain}},
  \bibinfo{journal}{Phys. Rev. B} \textbf{\bibinfo{volume}{58}},
  \bibinfo{pages}{R10167} (\bibinfo{year}{1998}),
  \urlprefix\url{http://link.aps.org/doi/10.1103/PhysRevB.58.R10167}.

\bibitem[{\citenamefont{T\ifmmode~\mbox{\H{o}}\else \H{o}\fi{}ke and
  Jain}(2006)}]{Toke2006}
\bibinfo{author}{\bibfnamefont{C.}~\bibnamefont{T\ifmmode~\mbox{\H{o}}\else
  \H{o}\fi{}ke}} \bibnamefont{and} \bibinfo{author}{\bibfnamefont{J.~K.}
  \bibnamefont{Jain}}, \bibinfo{journal}{Phys. Rev. Lett.}
  \textbf{\bibinfo{volume}{96}}, \bibinfo{pages}{246805}
  (\bibinfo{year}{2006}),
  \urlprefix\url{http://link.aps.org/doi/10.1103/PhysRevLett.96.246805}.

\bibitem[{\citenamefont{Haldane}(1983)}]{Haldane1983}
\bibinfo{author}{\bibfnamefont{F.~D.~M.} \bibnamefont{Haldane}},
  \bibinfo{journal}{Phys. Rev. Lett.} \textbf{\bibinfo{volume}{51}},
  \bibinfo{pages}{605} (\bibinfo{year}{1983}),
  \urlprefix\url{http://link.aps.org/doi/10.1103/PhysRevLett.51.605}.

\bibitem[{\citenamefont{{Prange} and {Girvin}}(1987)}]{Prange1987}
\bibinfo{author}{\bibfnamefont{R.~E.} \bibnamefont{{Prange}}} \bibnamefont{and}
  \bibinfo{author}{\bibfnamefont{S.~M.} \bibnamefont{{Girvin}}},
  \emph{\bibinfo{title}{The Quantum Hall Effect}}
  (\bibinfo{publisher}{Springer-Verlag}, \bibinfo{address}{New York},
  \bibinfo{year}{1987}).

\bibitem[{\citenamefont{Fano et~al.}(1986)\citenamefont{Fano, Ortolani, and
  Colombo}}]{Fano1986}
\bibinfo{author}{\bibfnamefont{G.}~\bibnamefont{Fano}},
  \bibinfo{author}{\bibfnamefont{F.}~\bibnamefont{Ortolani}}, \bibnamefont{and}
  \bibinfo{author}{\bibfnamefont{E.}~\bibnamefont{Colombo}},
  \bibinfo{journal}{Phys. Rev. B} \textbf{\bibinfo{volume}{34}},
  \bibinfo{pages}{2670} (\bibinfo{year}{1986}),
  \urlprefix\url{http://link.aps.org/doi/10.1103/PhysRevB.34.2670}.

\bibitem[{\citenamefont{Wu and Yang}(1976)}]{Wu1976}
\bibinfo{author}{\bibfnamefont{T.~T.} \bibnamefont{Wu}} \bibnamefont{and}
  \bibinfo{author}{\bibfnamefont{C.~N.} \bibnamefont{Yang}},
  \bibinfo{journal}{Nuclear Physics B} \textbf{\bibinfo{volume}{107}},
  \bibinfo{pages}{365 } (\bibinfo{year}{1976}), ISSN \bibinfo{issn}{0550-3213},
  \urlprefix\url{http://www.sciencedirect.com/science/article/pii/0550321376901437}.

\bibitem[{\citenamefont{Wu and Yang}(1977)}]{Wu1977}
\bibinfo{author}{\bibfnamefont{T.~T.} \bibnamefont{Wu}} \bibnamefont{and}
  \bibinfo{author}{\bibfnamefont{C.~N.} \bibnamefont{Yang}},
  \bibinfo{journal}{Phys. Rev. D} \textbf{\bibinfo{volume}{16}},
  \bibinfo{pages}{1018} (\bibinfo{year}{1977}),
  \urlprefix\url{http://link.aps.org/doi/10.1103/PhysRevD.16.1018}.

\bibitem[{\citenamefont{Scanio}(1977)}]{Scanio1977}
\bibinfo{author}{\bibfnamefont{J.~J.~G.} \bibnamefont{Scanio}},
  \bibinfo{journal}{American Journal of Physics} \textbf{\bibinfo{volume}{45}},
  \bibinfo{pages}{173} (\bibinfo{year}{1977}),
  \urlprefix\url{http://scitation.aip.org/content/aapt/journal/ajp/45/2/10.1119/1.10649}.

\bibitem[{\citenamefont{Dray}(1985)}]{Dray1985}
\bibinfo{author}{\bibfnamefont{T.}~\bibnamefont{Dray}},
  \bibinfo{journal}{Journal of Mathematical Physics}
  \textbf{\bibinfo{volume}{26}}, \bibinfo{pages}{1030} (\bibinfo{year}{1985}),
  \urlprefix\url{http://scitation.aip.org/content/aip/journal/jmp/26/5/10.1063/1.526533}.

\bibitem[{\citenamefont{Dray}(1986)}]{Dray1986}
\bibinfo{author}{\bibfnamefont{T.}~\bibnamefont{Dray}},
  \bibinfo{journal}{Journal of Mathematical Physics}
  \textbf{\bibinfo{volume}{27}}, \bibinfo{pages}{781} (\bibinfo{year}{1986}),
  \urlprefix\url{http://scitation.aip.org/content/aip/journal/jmp/27/3/10.1063/1.527183}.

\bibitem[{\citenamefont{Jain}(2007)}]{Jain2007}
\bibinfo{author}{\bibfnamefont{J.~K.} \bibnamefont{Jain}},
  \emph{\bibinfo{title}{Composite Fermions}} (\bibinfo{publisher}{Cambridge
  University Press}, \bibinfo{address}{Cambridge, UK}, \bibinfo{year}{2007}).

\bibitem[{\citenamefont{Ginocchio and Haxton}(1993)}]{Ginocchio1993}
\bibinfo{author}{\bibfnamefont{J.}~\bibnamefont{Ginocchio}} \bibnamefont{and}
  \bibinfo{author}{\bibfnamefont{W.}~\bibnamefont{Haxton}}, in
  \emph{\bibinfo{booktitle}{Symmetries in Science VI}}, edited by
  \bibinfo{editor}{\bibfnamefont{B.}~\bibnamefont{Gruber}}
  (\bibinfo{publisher}{Springer US}, \bibinfo{year}{1993}), pp.
  \bibinfo{pages}{263--273}, ISBN \bibinfo{isbn}{978-1-4899-1221-3},
  \urlprefix\url{http://dx.doi.org/10.1007/978-1-4899-1219-0_23}.

\bibitem[{\citenamefont{W\'ojs and Quinn}(1998)}]{Wojs1998}
\bibinfo{author}{\bibfnamefont{A.}~\bibnamefont{W\'ojs}} \bibnamefont{and}
  \bibinfo{author}{\bibfnamefont{J.~J.} \bibnamefont{Quinn}},
  \bibinfo{journal}{Solid State Communications} \textbf{\bibinfo{volume}{108}},
  \bibinfo{pages}{493 } (\bibinfo{year}{1998}), ISSN \bibinfo{issn}{0038-1098},
  \urlprefix\url{http://www.sciencedirect.com/science/article/pii/S0038109898003159}.

\bibitem[{\citenamefont{Quinn and W\'ojs}(2000)}]{Quinn2000}
\bibinfo{author}{\bibfnamefont{J.}~\bibnamefont{Quinn}} \bibnamefont{and}
  \bibinfo{author}{\bibfnamefont{A.}~\bibnamefont{W\'ojs}},
  \bibinfo{journal}{Physica E: Low-dimensional Systems and Nanostructures}
  \textbf{\bibinfo{volume}{6}}, \bibinfo{pages}{1 } (\bibinfo{year}{2000}),
  ISSN \bibinfo{issn}{1386-9477},
  \urlprefix\url{http://www.sciencedirect.com/science/article/pii/S1386947799000491}.

\bibitem[{\citenamefont{Johnson and Payne}(1991)}]{Johnson1991}
\bibinfo{author}{\bibfnamefont{N.~F.} \bibnamefont{Johnson}} \bibnamefont{and}
  \bibinfo{author}{\bibfnamefont{M.~C.} \bibnamefont{Payne}},
  \bibinfo{journal}{Phys. Rev. Lett.} \textbf{\bibinfo{volume}{67}},
  \bibinfo{pages}{1157} (\bibinfo{year}{1991}),
  \urlprefix\url{http://link.aps.org/doi/10.1103/PhysRevLett.67.1157}.

\bibitem[{\citenamefont{Johnson}(2002)}]{Johnson2002}
\bibinfo{author}{\bibfnamefont{B.~L.} \bibnamefont{Johnson}},
  \bibinfo{journal}{American Journal of Physics} \textbf{\bibinfo{volume}{70}},
  \bibinfo{pages}{401} (\bibinfo{year}{2002}),
  \urlprefix\url{http://scitation.aip.org/content/aapt/journal/ajp/70/4/10.1119/1.1446855}.

\bibitem[{\citenamefont{Edmonds}(1996)}]{Edmonds1996}
\bibinfo{author}{\bibfnamefont{A.~R.} \bibnamefont{Edmonds}},
  \emph{\bibinfo{title}{Angular Momentum in Quantum Mechanics}}
  (\bibinfo{publisher}{Princeton University Press},
  \bibinfo{address}{Princeton, New Jersey}, \bibinfo{year}{1996}).

\bibitem[{\citenamefont{Za\l{}uska-Kotur
  et~al.}(2000)\citenamefont{Za\l{}uska-Kotur, Gajda, Or\l{}owski, and
  Mostowski}}]{Zaluska-Kotur2000}
\bibinfo{author}{\bibfnamefont{M.~A.} \bibnamefont{Za\l{}uska-Kotur}},
  \bibinfo{author}{\bibfnamefont{M.}~\bibnamefont{Gajda}},
  \bibinfo{author}{\bibfnamefont{A.}~\bibnamefont{Or\l{}owski}},
  \bibnamefont{and}
  \bibinfo{author}{\bibfnamefont{J.}~\bibnamefont{Mostowski}},
  \bibinfo{journal}{Phys. Rev. A} \textbf{\bibinfo{volume}{61}},
  \bibinfo{pages}{033613} (\bibinfo{year}{2000}),
  \urlprefix\url{http://link.aps.org/doi/10.1103/PhysRevA.61.033613}.

\bibitem[{\citenamefont{{Quinn} and {Yi}}(2009)}]{Quinn2009}
\bibinfo{author}{\bibfnamefont{J.~J.} \bibnamefont{{Quinn}}} \bibnamefont{and}
  \bibinfo{author}{\bibfnamefont{K.-S.} \bibnamefont{{Yi}}},
  \emph{\bibinfo{title}{Solid State Physics, Principles and Modern
  Applications}} (\bibinfo{publisher}{Springer-Verlag}, \bibinfo{address}{New
  York}, \bibinfo{year}{2009}).

\bibitem[{\citenamefont{U.~Fano}(1959)}]{Fano1959}
\bibinfo{author}{\bibfnamefont{G.~R.} \bibnamefont{U.~Fano}},
  \emph{\bibinfo{title}{Irreducible Tensorial Sets}}
  (\bibinfo{publisher}{Academic Press}, \bibinfo{address}{New York},
  \bibinfo{year}{1959}).

\end{thebibliography}

\end{document}